\documentclass[aps,physrev,preprint,groupedaddress]{revtex4-2}

\usepackage{amsmath}
\usepackage{graphicx}
\begin{document}
\title{Spectrotemporal processing in a dual gradient echo and electromagnetically-induced transparency memory}

\author{Jesse L Everett}
\affiliation{Research School of Physics\\ Australian National University}


\begin{abstract}
Spectrotemporal encoding of optical quantum information is emerging as a powerful tool in quantum information technology. Processing of spectrotemporal information has recently been demonstrated in multi-mode quantum memories, based on extensions to memory protocols. We simulate one such process, the fractional Fourier transform, in a system based on a dual quantum memory composed of successive gradient echo memory and electromagnetically-induced transparency operations. We demonstrate the potential of electromagnetically-induced transparency systems for spectrotemporal processing.
\end{abstract}

\maketitle
\thispagestyle{empty}

\section*{Introduction}
Processing of optical quantum signals in multi-mode ensemble memories is a well established technique that forms the basis for a range of applications in quantum information processing. These applications include generating quantum signals \cite{hellerColdAtomTemporallyMultiplexed2020,dideriksenRoomtemperatureSinglephotonSource2021}, heralding or entangling non-deterministic photon sources \cite{tsaiQuantumStorageManipulation2020,liuHeraldedEntanglementDistribution2021,davidsonSinglePhotonSynchronizationRoomTemperature2023,leungHighlyEfficientStorage2024a}, frequency conversion \cite{bustardQuantumFrequencyConversion2017,radnaevQuantumMemoryTelecomwavelength2010a}, multiplexing \cite{sinclairSpectralMultiplexingScalable2014a,hellerColdAtomTemporallyMultiplexed2020,tellerSolidstateTemporallyMultiplexed2025}, timing, bandwidth manipulation and optical RAM \cite{hosseiniStorageManipulationLight2012}, and intra-memory linear computing operations \cite{campbellConfigurableUnitaryTransformations2014b}). 

The power of quantum information processing increases with the dimension of the information. Spatial degrees of freedom such as transverse modes and orbital angular momentum have been used in communication \cite{xavierQuantumInformationProcessing2020} and quantum information encoding \cite{erhardTwistedPhotonsNew2018a,erhardAdvancesHighdimensionalQuantum2020}. Higher-dimensional information can also be encoded in the time-frequency domain, and subsets of the domain such as time-bin and frequency-bin encoding are relevant to many of the above examples. Spectrotemporal encodings such as the Hermite-Gauss modes \cite{brechtPhotonTemporalModes2015a} have many attractive properties for encoding quantum information \cite{karpinskiControlMeasurementQuantum2021}. However, these modes overlap in both time and frequency and so require processes that affect spectral and temporal properties simultaneously - 'spectrotemporal manipulation'.

 Techniques developed for manipulating temporal modes of ultrashort pulses  are analogous to imaging techniques for manipulating spatial modes, with dispersion standing in for spatial propagation and signal chirping (time-lens) replacing a spatial lens \cite{kolnerSpacetimeDualityTheory1994a,sosnickiInterfacePicosecondNanosecond2023}. The addition of time-dependent phase shifts completes the toolkit \cite{joshiPicosecondresolutionSinglephotonTime2022,horoshkoInterferometricSortingTemporal2024}. Spectrotemporal manipulation has also been demonstrated with the gradient echo memory (GEM) \cite{sparkesPrecisionSpectralManipulation2012, nieweltExperimentalImplementationOptical2023, mazelanikOpticaldomainSpectralSuperresolution2022a}. Implementations in these two types of systems rely on different physical effects due to the narrow bandwidth of ensemble memories.
 
Electromagnetically-induced transparency (EIT) is another technique for high efficiency optical quantum memory \cite{novikovaElectromagneticallyInducedTransparencybased2012,hsiaoHighlyEfficientCoherent2018a}, and is also used in sensing \cite{finkelsteinPracticalGuideElectromagnetically2023}. The EIT mechanism is less advantageous for spectral manipulation in ensembles due to its limited bandwidth, though a dual GEM-EIT memory was recently used to implement a time-frequency Fourier transform (FT) \cite{papnejaDemonstrationPhotonicTimefrequency2025a,kurzynaHybridQuantumMemory2025}. 

 Here, a further application in spectrotemporal manipulation for EIT is demonstrated through the extension of the dual memory Fourier transform to a fractional Fourier transform (FrFT). The FrFT is a generalisation of the FT as an operator that includes a fractional power \cite{namiasFractionalOrderFourier1980}. The FT can be considered a rotation by $\pi/2$ in the 2D time-frequency phase-space of time and frequency, which exchanges the time and frequency information in the signal. The FrFT is then an arbitrary rotation in the phase space, where the fractional power sets the rotation angle relative to $\pi/2$. 

We implement the FrFT in a simulated EIT-GEM system by chirping the memory frequency during storage and recall. This is similar to the technique used with GEM to implement the FrFT \cite{nieweltExperimentalImplementationOptical2023}, where Niewelt et al. used a 3-step process of chirping, applying dispersion, then chirping \cite{nieweltExperimentalImplementationOptical2023}. Chirp and dispersion are orthogonal shears in the time-frequency phase space, so the sequence amounts to a rotation by shearing. The EIT-GEM dual memory protocol removes the explicit application of dispersion, eliminating the requirement for spatial control of the memory. 

The relatively simple spectrotemporal manipulation of the FrFT could be extended by concatenating memories to apply multiple transformation steps for operations such as mode sorting \cite{joshiPicosecondresolutionSinglephotonTime2022}.

\section{Theory}
The fractional Fourier transform is defined \cite{namiasFractionalOrderFourier1980} as 
\begin{align}
    \mathcal{F}_\alpha f(x) =\frac{\exp{i(\frac{\pi-2\alpha}{4})}}{\sqrt{2\pi\sin\alpha}}\exp({-\frac{i x'^2}{2}\cot{\alpha} })\int_{-\infty}^{+\infty}\exp\left(-\frac{ix'^2}{2}\cot\alpha+\frac{ixx'}{\sin\alpha}\right)f(x')dx'.
    \end{align} Setting $\alpha=\pi/2$ reduces this to a Fourier transform. 

It has been established by analogy to 2-dimensional image rotations that the FrFT can be implemented by 3 shears in time-frequency phase space performed within a multi-mode optical memory  \cite{nieweltExperimentalImplementationOptical2023}. An image rotation can also be decomposed into 2 orthogonal shears and a scaling, and we show how to apply this in the GEM-EIT system to produce a FrFT.

\subsection{Idealised scheme} 
\begin{figure}
    \centering
    \includegraphics[width=0.75\linewidth]{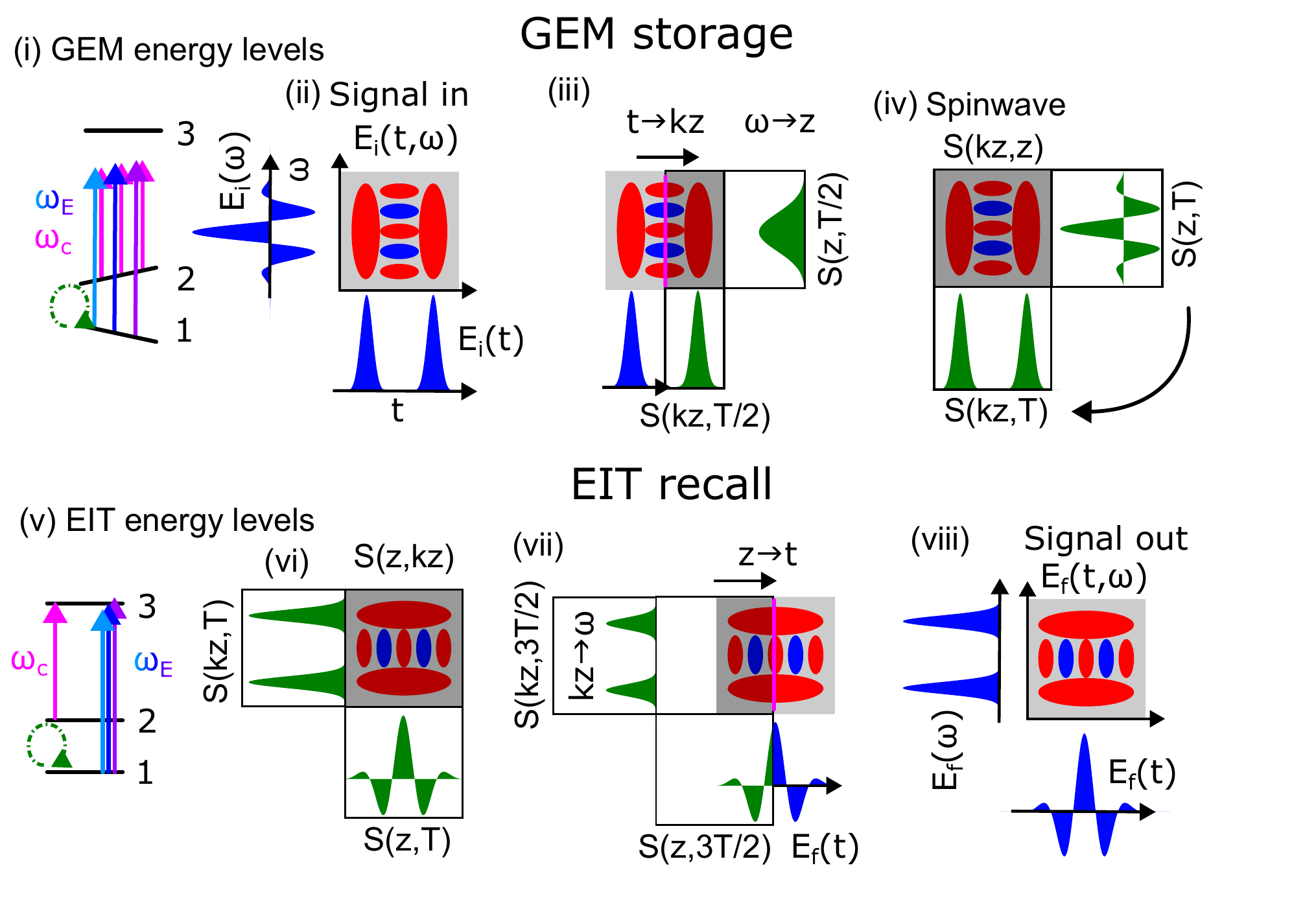}
    \caption{Conceptual implementation of the Fourier transform: (i) The GEM level scheme, where different frequencies of the input signal (shades of blue) are two-photon resonant with the control laser (pink) at different positions, producing a spinwave (dashed, green). (ii) Time and frequency components and WDF of the signal. WDF has positive (red) and negative (blue) regions. (iii) Partial storage; the first pulse is converted to spinwave (green).  The gradient transports the spinwave in momentum ($k_z$).  (iv) Complete storage; the second pulse is converted to spinwave, overlapping spatially and interfering with the first pulse to give a spatially modulated spinwave ($S(z,T)$). (v) EIT level scheme; the resonant control laser produces slow light and transports the spinwave in $z$. (vi) The spinwave WDF rotated by $-\pi/2$ to indicate the new transport dimension. (vii) Partial recall; spinwave reaches the end of the ensemble and is converted entirely to light (interface marked by pink line); the spatial envelope of the spinwave is converted to the temporal envelope of the output signal, the momentum distribution becomes the output spectrum. (viii) Complete recall; The output signal is the time-frequency Fourier transform of the input.}
    \label{fig:concepta}
\end{figure}

\begin{figure}
    \centering
    \includegraphics[width=0.90\linewidth]{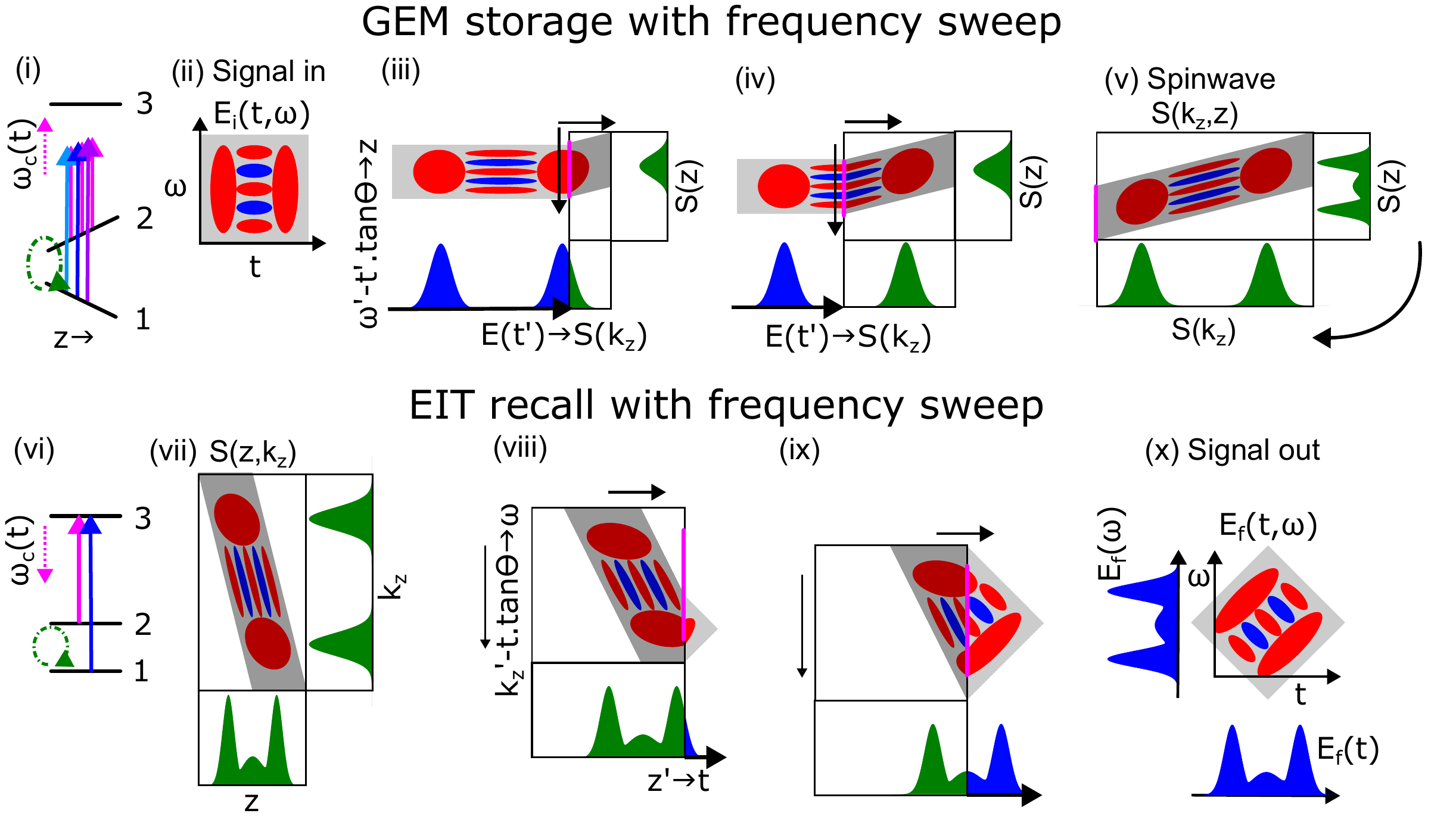}
    \caption{Extension of the conceptual diagram to a fractional Fourier transform: (i) Increased gradient to accommodate frequency sweep, which follows the dotted pink arrow. (iii) The signal occupies a smaller part of the memory bandwidth, and the signal WDF is rescaled to $(t',\omega')$ to preserve the scale of $(k_z,z)$. Sweeping the control frequency changes the position of the interface in $z$ as a function of time, producing a sheared spinwave WDF. (v) The stored spinwave. The spatial envelope is smeared due to the shear, while the momentum distribution is only stretched (vii-viii) The EIT velocity, $z\rightarrow t$ is modified to reverse the distortion from 2-shear rotation. The rescaled $z'$ sets the spinwave WDF at the same width as the rotated output signal. (x) The output signal, rotated $-\pi/4$  in time-frequency.}
    \label{fig:conceptb}
\end{figure}

The FT in GEM-EIT and the extension to the FrFT are shown in the concept diagrams in Figs. \ref{fig:concepta} and \ref{fig:conceptb}. In both these memories, a controllable coherent interaction between light and the atomic ensemble converts the optical signal to a collective atomic excitation - a spinwave - and converts the spinwave back to an optical signal. The interactions rely a on 2-photon resonance with a bright control laser, and the frequency and intensity of the control can be used to change the frequency and bandwidth of the memory. For this explanation, an exact mapping from the time and frequency of the signal to the position and momentum of the spinwave is assumed, although the physical systems have a more complicated mapping \cite{hushAnalysisOperationGradient2013, everettElectromagneticallyinducedTransparencyAssists2024, moiseevRephasingProcessesQuantum2012}. The distribution of a signal in the time-frequency phase space, or a spinwave in the position-momentum phase space, can be visualised with a Wigner Distribution Function (WDF) \cite{wignerQuantumCorrectionThermodynamic1932,bastiaansWignerDistributionFunction1978}. A cartoon WDF is used to show how the phase-space rotation due to the dual encoding, chirps, and rescaling, implements the FT and FrFT.

The FT is performed by storing with GEM and retrieving with EIT. During GEM storage, a longitudinal frequency gradient in the atomic absorption maps the frequency information in the optical signal into the longitudinal position $z$ of the spinwave. The frequency gradient winds up the phase of the spinwave, transporting it through momentum space, shown in Fig. \ref{fig:concepta} a)(ii), resulting in the arrival time being encoded as momentum. The same spinwave can then be recalled with EIT, where the slow light effect transports the spinwave along the spatial dimension of the memory (horizontal, after rotating the spinwave WDF by $-\pi/2$). The spinwave converts entirely to optical signal at the far end, with the output signal time corresponding to the original spinwave position, and the frequency corresponding to spinwave momentum. The rotation of the spinwave WDF between \ref{fig:concepta} a)(iii) and (iv) indicates the exchange of the transport dimension, and the pink line in Fig. \ref{fig:concepta} a)(v) shows the conversion of the spinwave spatial position to signal recall time. The time and frequency components are obtained from the WDF by projection (integrating over the other dimension), so the rotation by $\pi/2$ of the WDF represents the exchange of those domains in the signal.

The FT is extended to a FrFT in Fig. \ref{fig:conceptb}. The GEM bandwidth is increased, memory frequency chirps are added during storage and recall, and the EIT recall speed is scaled, rotating the input sgnal by $\theta=-\pi/4$. The first chirp, applied by sweeping the control frequency while storing with GEM, changes the position at which the signal is stored into the spinwave as a function of time. The pulses arriving at different times are stored in different positions, and the spinwave WDF is sheared in $z$. The memory bandwidth is increased by the size of the chirp so that the signal is still entirely stored within the memory. The increased gradient also increases the momentum transport speed. The scaling of the ensemble length $z$ is kept mostly constant in the diagram as a reference for the remaining dimensions, allowing the trigonometric derivation of the required shears. By properly scaling between signal and spinwave, the spinwave can be correctly transformed to achieve the desired final signal. The mapping from optical signal time to spinwave momentum $t'\rightarrow k_z$ and from signal frequency to spinwave position $\omega'\rightarrow z$, is shown from Fig. \ref{fig:conceptb} (ii-iii). The stored, sheared spinwave is shown in Fig. \ref{fig:conceptb} (v). The WDF projection onto $k_z$ is unaffected by the shear, but in $z$ the spinwave envelope is modified, as expected by the moving storage location. The spinwave WDF is again rotated by $-\pi/2$ for EIT decoding $z\rightarrow t$. The chirp during the EIT recall, again applied by sweeping the control frequency, changes the frequency of the output signal as a function of time, represented by a vertical motion of the spinwave from (vii-ix). Once the signal is recalled its frequency does not change, so the chirp shears the output signal WDF in $\omega$. The swapping of the encoding and decoding dimension causes the two shears to become orthogonal. Rotation by only two shears requires an additional scaling, which is applied by reducing the EIT recall speed (Fig. \ref{fig:conceptb} (v)-(vi)). After this scaling, the second shear produces the desired $-\pi/4$ rotation on the output signal.

The conceptual diagram was used to derive the protocol for simulating FrFT. The input bandwidth $W_i$ and duration $T_i$ set the boundaries of the spinwave and signal WDF. The required chirps were then trigonometrically derived by considering the rotation of a square, and the resulting angles of the sides and positions of the corners. This gives an input chirp of $\partial_t\omega=-(W_i/T_i)\tan\theta_{FrFT}$, calculated from the position of the square phase space's corners after the rotation. The chirp is applied by sweeping the control field frequency, wich changes the location of 2-photon resonance. The storage bandwidth is increased to $(1+|\tan\theta_{FrFT}|)W_i$. This produces the same sheared, rescaled spinwave as in the conceptual diagram. The recall time is set to $T_f=(|\sin\theta_{FrFT}|+|\cos\theta_{FrFT}|)T_i$, the width of the final rotated square (the corners stick out). The passage of the spinwave across $z$ has to last this amount of time, so the EIT group velocity is adjusted to $v_g=L/T_f$. Finally, the output chirp is also $\partial_t\omega=-W_i\tan\theta_{FrFT}$, calculated from the position of the remaining corners.

Due to the $\tan\theta$ dependence of the chirps, the FrFT rotations are limited to $\theta< \pi/2$. However, the FrFT rotation is also added to the $\pm\pi/2$ rotation from the FT, so any rotation (except approaching 0 or $\pi$) is possible. The sign of the FT can be changed by swapping the sign of the GEM storage gradient. 

\subsection{Numerical Modelling}
In a physical system the mapping from the signal to the spinwave, $\mathcal{E}(t,\omega) \rightarrow S(z,k_z)$ is complicated. Dispersion within the atomic medium and due to the two-photon interactions distorts the stored spinwave and, accordingly, the output signal \cite{hushAnalysisOperationGradient2013, hosseiniStorageManipulationLight2012,everettElectromagneticallyinducedTransparencyAssists2024}. Fine tuning of the protocol, particularly the chirp magnitudes, will be necessary. Further, the two memory protocols have different effective bandwidths and efficiencies, so modelling is essential to understanding the capabilities of the system for this transformation. In the experimental demonstration of GEM-EIT, simulations based on numerical solution of the optical Maxwell-Bloch equations could accurately describe the system \cite{papnejaDemonstrationPhotonicTimefrequency2025a}, so those simulations are adapted by adding the frequency chirps for the FrFT. Predictions of fidelity and efficiency are compared with the simulated results.

The following equations describing a 3-level semiclassical approximation suitable for ensemble memories \cite{gorshkovPhotonStorageType2007c} were numerically integrated in XMDS2 \cite{dennisXMDS2FastScalable2013}.
\begin{align}
    \partial_t S &= i\Omega^* P - (\gamma_S + i z\delta_{grad}(t)+i\delta_{chirp}(t))S\\
    \partial_t P& = i\sqrt{d}\gamma\mathcal{E} + i\Omega S - (\gamma + i\Delta) P\\
    \partial_z \mathcal{E} &= i\sqrt{d} P\label{dz}
\end{align}
with a spin coherence $S$, optical coherence $P$, optical field envelope $\mathcal{E}$, excited state decay rate $2\gamma$, optical depth $2d$, control Rabi frequency $\Omega/2$, control detuning $\Delta$, and two-photon detuning for gradient $\delta_{grad}$ and chirp $\delta_{chirp}$, and a normalised length $L=1$. 

An ensemble of rubidium-87 atoms interacting on the D1 line at an experimentally achievable optical depth of 1000 \cite{tranterMultiparameterOptimisationMagnetooptical2018a} provides the constant parameters for the simulation. GEM storage is set for duration $T=10 \mu$ s$, \Delta=250$ MHz, and a gradient $\delta_{grad}$. The detuning is set to $\Delta=0$EIT group velocity for recall is set using $\Omega_{EIT}$ for a recall duration  $(|\cos\theta|+|\sin\theta|)T$.

The Hermite-Gauss modes $HG_n$ are useful test signals for this simulation. The $HG_n$ are eigenfunctions of the FrFT \cite{namiasFractionalOrderFourier1980}, acquiring only a phase proportional to the rotation and mode number, $\phi=n\theta$. To measure different time-bandwidth products using these modes while keeping the simulation timing constant, a 'mode volume' $m$ was set by changing the $HG$ bandwidth, so that modes up to $HG_m$ fit within the storage duration. The GEM bandwidth was increased accordingly to maintain the $HG^m_n$ as eigenmodes (the $HG^m$ WDF is circular at that bandwidth, and the FrFT rotation leaves the WDF constant except for a phase). Modes $HG^{m=10}_n$, for $0\leq n\leq10$ were stored and transformed with $FrFT^m(\theta)$ for $\pi/12\leq\theta\leq11\pi/12$. 

Applied chirps for both GEM and EIT are physically possible within the chosen parameters by sweeping the control field frequency. The maximum chirp should be comparable to the atomic linewidth to ensure the EIT dynamics remain constant \cite{novikovaElectromagneticallyInducedTransparencybased2012}. For the largest chirp, with $m=10, \theta=\pi/12$, this is $\approx2\gamma\times2\pi$ - almost exactly the rubidium-87 D1 linewidth. For larger bandwidths, the one-photon detuning $\Delta$ could be reduced in EIT by sweeping the magnetic field along with the control field frequency. The GEM interaction remains consistent over a much larger control field frequency range, so this chirp is not the limiting factor.

\section{Results and Discussion}

\begin{figure}
    \centering
    \includegraphics[width=0.6\linewidth]{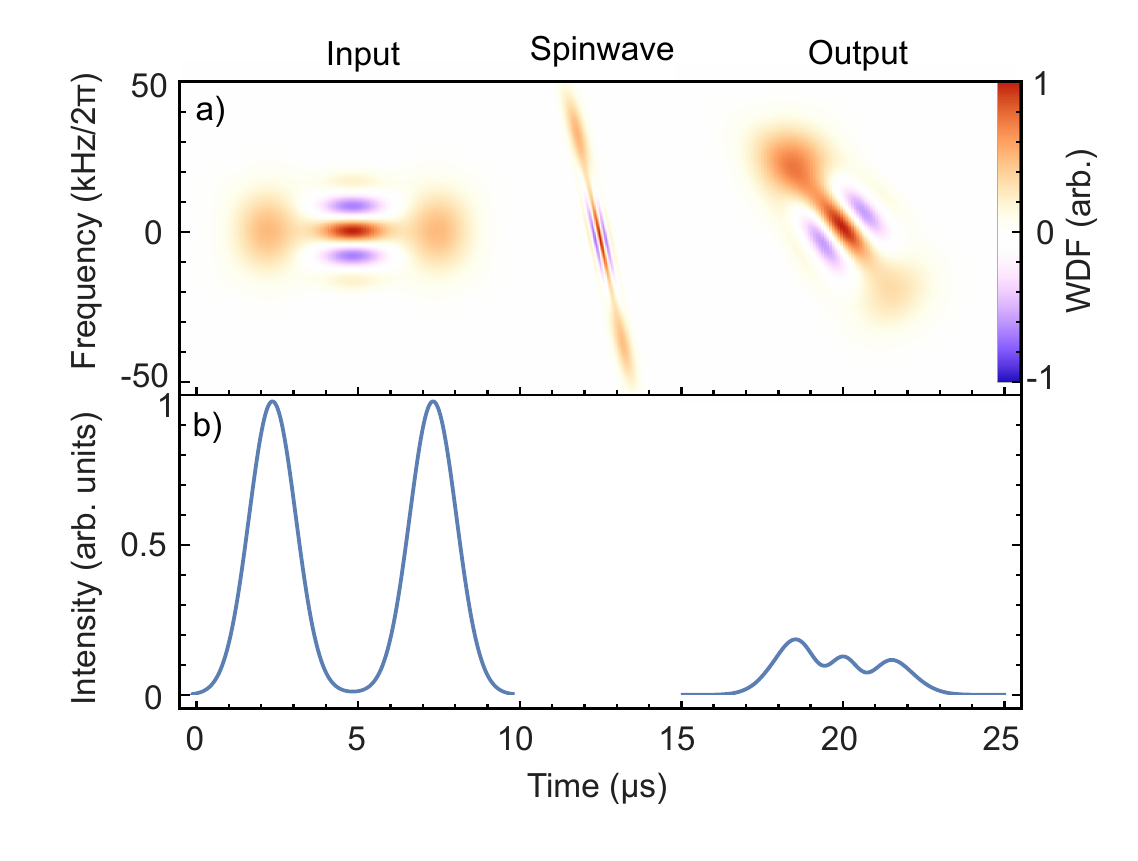}
    \caption{Simulation of a rotation of two Gaussian pulses by $-\pi/4$, as in the concept diagram. a) WDF of the input, stored spinwave, and outputs. The WDF amplitudes are not to scale. b) Intensity of the input and output signals.}
    \label{fig:example}
\end{figure}

Fig. \ref{fig:example} shows an example simulation result. The spinwave WDF plotted is scaled similar to that in Fig. \ref{fig:conceptb} (vii). The WDF is calculated from the input and output pulses and the stored spinwave, according to
\begin{align*}
    F(t,f)=\int{E(t+\tau/2)E^*(t-\tau/2)e^{-2\pi i ft}\mathrm{d}\tau},\\
    F(z,k_z)=\int{E(z+\xi/2)E^*(z-\xi/2)e^{ i zk_z}\mathrm{d}\xi}
\end{align*},
and the plots are scaled for a square aspect ratio.

\begin{figure}
    \centering
    \includegraphics[width=\linewidth]{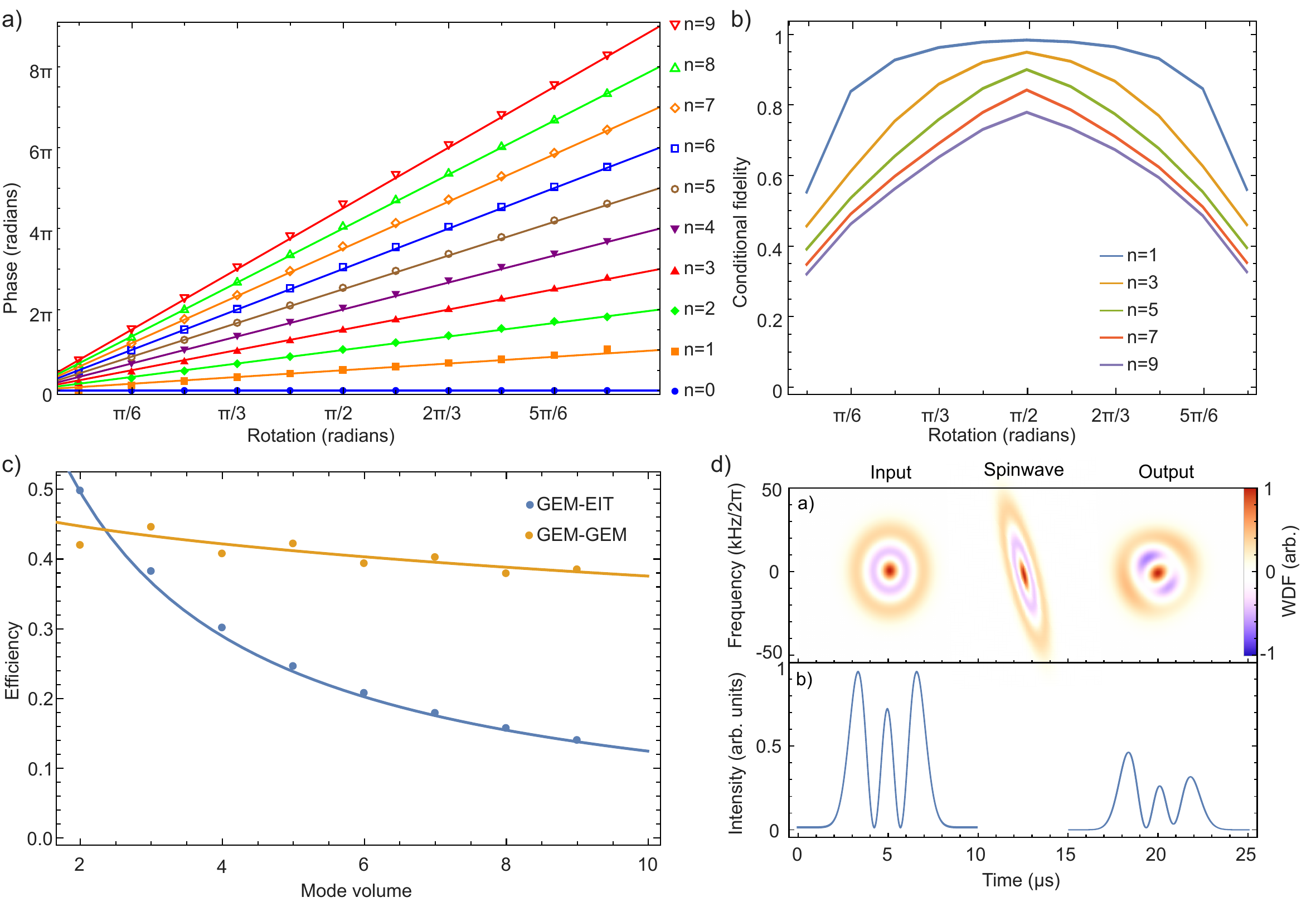}
    \caption{Simulation results. a) Eigenphase measurement of the Hermite-Gauss modes for $HG_n$ and $m=10$. Lines show the expected phase for each mode vs rotation. b) Conditional fidelity for selected HG modes and rotations at $m=10$. c) Efficiency of a $\pi/4$ rotation in GEM-EIT and GEM-GEM for $HG_m$ for $m<10$. d) Example of a $HG_2$ mode with $FrFT^{m=10}(-\pi/4)$ rotation.}
    \label{fig:results}
\end{figure}

The full set of simulations are used to show the accuracy and fidelity of the FrFT, as well as the efficiency with increasing time-bandwidth product. Fig. \ref{fig:results}  a) shows the eigenphase measurements taken by multiplying the output signal with the input and integrating over time, confirming that the output phases closely match the predicted $n\theta$.  Fig.\ref{fig:results} b) shows the conditional fidelity for several of the HG modes for the various rotations, where the conditional fidelity is defined as the overlap of the output pulse with the desired output, divided by the efficiency. The efficiency is the integrated output intensity divided by the integrated input intensity. Increased rotation lowers the conditional fidelity, since the higher frequency components are attenuated more strongly during EIT recall. The higher modes have more high frequency components, so also have lower conditional fidelity.

Fig. \ref{fig:results} c) compares the efficiency of the GEM-EIT two-shear protocol with a GEM-GEM three-step protocol similar to that of Niewelt et al. \cite{nieweltExperimentalImplementationOptical2023}. Fitted lines are added that follow the theoretical efficiency scaling of optical depth per bandwidth. The $HG_m$  signal bandwidth and memory bandwidth $\eta$ increase roughly as $\sqrt{m}$, and the GEM control field $\Omega_{GEM}$ must increase as $\sqrt{\eta}$ to maintain the interaction strength. Control field scattering causes exponential loss scaling with $|\Omega_{GEM}|^2$, and GEM efficiency
 therefore has a roughly exponential scaling of $\exp{(-\sqrt{m})}$. The scaling of GEM-EIT efficiency with $m$ is more severe, because the loss is caused by incoherent absorption of the signal spectrum outside the EIT transparency window. Roughly $1/m$ of the $HG^m$ signal is within the window for rotations close to $\pi/4$m and the high optical depth causes most of the signal outside the transparency window is lost. This gives a ~$1/m$ scaling for the efficiency. The EIT efficiency scales more strongly with mode volume than GEM, so there is a large gap between EIT-GEM and GEM-GEM efficiencies. Higher optical depths correspond to larger transparency windows, allowing more efficient EIT-GEM FrFT. 
 
 Finally, the oscillating efficiency of the GEM-GEM transformation is caused by interrupted interference between the incoming signal and the already-stored signal. Normally, this would be constructive, enhancing the storage of the incoming signal. However, the signal contains lobes with opposite phase, and the motion of the storage region with the chirp brings these into contact with the incoming signal. This effect can be seen in Fig. \ref{fig:results} d), where some of the negative phase part of the spinwave and output are reduced. Reducing the interaction strength can reduce this interference, and the size of the effect, at the cost of reducing the efficiency. 

We have analysed the operation of a fractional Fourier transform based on the dual memory protocol EIT-GEM and shown that the dual memory can be used for this purpose, allowing spectrotemporal manipulation without requiring additional spatial access to the spinwave. More complex spectral manipulations such as mode sorting and multiplexing would also be possible by time-varying chirps and recall bandwidths, and by concatenating memories for multi-stage operations. Quantum sensing and spectroscopic applications of EIT might also benefit from the technique, if a chirped EIT interaction can be combined with a sufficient dispersion to allow a controllable rotation of the measurement phase-space.

\section{Acknowledgements}
This work was funded by the Australian Research Council Centre of Excellence for Quantum Computation and Communication Technology (CE170100012).

The author thanks Ben Buchler and Cameron Trainor for valuable discussions.

\bibliography{frftbib}

\end{document}